\documentclass[aps, pre, twocolumn, groupedaddress, superscriptaddress, amsmath, amssymb, floatfix,showpacs]{revtex4-1}

\usepackage{graphicx}
\usepackage{dcolumn}
\usepackage{bm}
\usepackage{placeins}
\usepackage{graphicx,color}
\usepackage{caption}
\usepackage{subcaption}
\usepackage{ragged2e}

\begin{document}

\preprint{APS/123-QED}

\title{Black-Hole Ringdown from Photon-Sphere Modes in an Elastic Analogue}

\author{Jiaoqing Wang}
\affiliation{
Department of Physics, The Jack and Pearl Resnick Institute for Advanced Technology, Bar-Ilan University, Ramat-Gan, 5290002 Israel}

\author{Chenni Xu}
\affiliation{
Department of Physics, The Jack and Pearl Resnick Institute for Advanced Technology, Bar-Ilan University, Ramat-Gan, 5290002 Israel}

\author{Sophia Purow}
\affiliation{
Department of Physics, The Jack and Pearl Resnick Institute for Advanced Technology, Bar-Ilan University, Ramat-Gan, 5290002 Israel}

\author{Sébastien Guenneau}
\affiliation{
Department of Physics, Imperial College London, London SW7 2AZ, United Kingdom}
\affiliation{
CNRS–Imperial Abraham de Moivre International Research Laboratory, Imperial College London, London, SW7 2AZ, United Kingdom}

\author{Patrick Sebbah}
\thanks{Corresponding author: patrick.sebbah@biu.ac.il}
\affiliation{
Department of Physics, The Jack and Pearl Resnick Institute for Advanced Technology, Bar-Ilan University, Ramat-Gan, 5290002 Israel}
\begin{abstract}

Quasinormal modes (QNMs) govern the ringdown of black holes, yet their physical origin and experimental accessibility remain limited. Here we create an elastic analogue of a black hole using a curved metallic shell engineered to emulate a Schwarzschild-like metric, and directly probe its vibrational dynamics in the time domain. Following impulsive excitation, the elastic system exhibits a clear ringdown response, providing a tangible analogue of black-hole ringing. The underlying modes are flexural vibrations localized near the photon sphere. A reduction of the thin-shell elastic equations yields a photon-sphere effective potential and predicts the dependence of the localized-mode frequencies on azimuthal order. Crucially, they persist even when boundary reflections are strongly suppressed, demonstrating that their confinement is governed by a curvature-induced effective potential rather than by cavity boundaries. This platform enables direct access to quasinormal modes through both their temporal decay and spectral signatures, establishing a direct connection between photon-sphere dynamics and ringdown. More broadly, our results identify curvature-induced trapping as a general mechanism for wave localization and dissipation in open systems.

\end{abstract}

\maketitle

\section{Introduction}
Black-hole mergers release intense bursts of gravitational radiation, now routinely detected by ground-based interferometers such as LIGO and Virgo \cite{portegies2000black,abbott2016tests,kagra2020prospects}. The final stage of this process, known as the ringdown, consists of damped oscillations that encode the properties of the final black hole. These signals are described in terms of quasinormal modes (QNMs), the natural response of spacetime to perturbations, whose frequencies are closely tied to the photon sphere: the region outside the event horizon where light can orbit the black hole. In this sense, a perturbed black hole behaves analogously to a struck bell, emitting a characteristic spectrum while dissipating energy over time. While the frequencies of these modes are now routinely extracted from observations, their physical nature---how they emerge, how they decay, and what governs their spatial structure---remains only partially understood, and is the subject of active research in the emerging field of black-hole spectroscopy.

Accessing this dynamical behavior in a controlled setting remains challenging. Analogue-gravity platforms have been developed to emulate aspects of curved spacetime in laboratory systems, following the seminal proposal by Unruh \cite{unruh1981experimental}. A wide range of implementations has since been explored, including fluids and Bose–Einstein condensates \cite{lahav2010realization,toshmatov2023phonon}, optical systems \cite{xu2021theory,ba2022conformal,trappingphotons}, and more recently curved photonic cavities supporting photon-sphere modes \cite{xu2026photon}. In these systems, quasinormal modes are typically inferred from steady-state or gain-selected responses, which provide access to their spectral signatures and spatial profiles. However, this approach does not directly probe the intrinsic temporal relaxation associated with ringdown, nor does it fully resolve the relation between mode selection, dissipation, and the underlying confinement mechanism. As a result, the physical connection between photon-sphere localization and quasinormal ringing remains only indirectly established.

In parallel, elastic waves in structured solids have emerged as a robust and highly tunable platform for investigating wave dynamics in engineered geometries \cite{krylov2014acoustic,lefebvre2015experiments}, as illustrated by the experimental realization of a wormhole analogue using transformation elastodynamics \cite{zhu2018elastic}. Beyond their versatility, elastic systems offer a particularly intuitive analogue of black-hole ringdown: following an impulsive excitation, they naturally exhibit damped vibrational responses closely resembling the ringing of perturbed compact objects. 

Here, we exploit this analogy by realizing an elastic black-hole analogue based on a curved metallic shell engineered to emulate a Schwarzschild-like metric. We derive an effective-potential description directly from
the thin-shell elastic equations, connecting the mapped geometry to both the spatial confinement and characteristic frequencies of the flexural modes. Using laser vibrometry and numerical simulations, we directly observe the temporal ringdown and spatial structure of flexural modes localized near the photon sphere. By introducing strong boundary absorption, we further demonstrate that these modes remain confined even when boundary reflections are suppressed, establishing that their localization originates from a curvature-induced effective potential rather than from cavity boundaries.

\section{Quasinormal Modes of Black Holes and Elastic Shells}

Black-hole perturbations admit an effective-potential description that provides the natural starting point for the elastic-shell analogue. We consider a static, spherically symmetric spacetime written in Schwarzschild-like coordinates $(t,\rho,\theta,\phi)$ as
\begin{equation}
ds^2=-f(\rho)c^2dt^2+f^{-1}(\rho)d\rho^2+\rho^2\left(d\theta^2+\sin^2\theta\,d\phi^2\right),
\label{eq:metric}
\end{equation}
where $\rho$ is the areal radial coordinate, $c$ is the speed of light, and $f(\rho)$ specifies the black-hole geometry. For a Schwarzschild black hole, $f(\rho)=1-r_s/\rho$, with $r_s=2GM/c^2$ the Schwarzschild radius, $G$ the gravitational constant, and $M$ the black-hole mass.

Perturbations of a static, spherically symmetric black hole can be reduced to a one-dimensional radial wave equation in an effective potential,
\begin{equation}
\frac{d^2\Psi}{d\rho_*^2}
+\left[\omega^2 - V_{\rm eff}(\rho)\right]\Psi=0 .
\label{eq:radial_wave}
\end{equation}
Here $\omega$ is the quasinormal-mode frequency and $\rho_*$ is the tortoise coordinate, defined by $d\rho_*=f^{-1}(\rho)d\rho$. This equation has the form of a stationary Schr\"odinger equation, with $\omega^2$ playing the role of an energy and $V_{\rm eff}$ that of a scattering potential. The potential is written as a function of the areal radius $\rho$, with the dependence on $\rho_*$ implicitly understood through $\rho=\rho(\rho_*)$.

In the eikonal limit, where the angular momentum is large, the dominant part of this potential is governed by the null-geodesic dynamics (Konoplya and Zhidenko, 2023 \cite{konoplya2023analytic}). It can be written, up to an irrelevant normalization, as
\begin{equation}
V_{\rm eff}(\rho)=\frac{L^2}{2\rho^2}f(\rho),
\label{eq:effective_potential}
\end{equation}
where $L$ is the conserved angular momentum of the corresponding null orbit. The photon sphere is located at an extremum of this potential, $dV_{\rm eff}/d\rho=0$, and its stability is determined by the sign of $d^2V_{\rm eff}/d\rho^2$.

In this regime, the connection between photon-sphere dynamics and quasinormal ringing becomes explicit \cite{ferrari1984new,konoplya2003quasinormal,cardoso2009geodesic}. The real part of the QNM frequency is set by the orbital frequency of waves circulating near the photon sphere, whereas the imaginary part is governed by their leakage rate, or equivalently by the Lyapunov exponent of the orbit. QNMs are therefore resonances of an open wave system rather than normal modes of a closed cavity.
This viewpoint suggests a direct mechanical implementation: a thin elastic shell behaves as a bell-like open resonator, supporting flexural vibrations that ring down through radiation and material losses. 

By engineering the shell geometry, one can shape an effective
potential for flexural vibrations and reproduce the spatial
confinement associated with photon-sphere dynamics. We consider
a thin shell of revolution with a slowly varying midsurface
radius $R(x)$, where $x$ is the meridional arclength. Within
the local cylindrical approximation to the
Donnell--Mushtari--Vlasov equations
\cite{saijyou2002analysis,soedel2005vibrations},
the out-of-plane displacement is written as
\begin{equation}
w(x,\phi)=\Phi(x)\cos(m\phi),
\end{equation}
where $m$ is the azimuthal mode number and the time dependence
$e^{-i\omega t}$ is understood.

Expanding the local dispersion relation about the flexural
branch near the circular-orbit limit yields, to leading order,
a one-dimensional wave equation along the shell profile,
\begin{equation}
\frac{d^2\Phi}{dx^2}
+\left[k_{\rm b}^2(\omega)-V_{\rm curv}(x)\right]\Phi=0,
\label{eq:elastic_radial_wave}
\end{equation}
with
\begin{equation}
V_{\rm curv}(x)=\frac{m^2}{R^2(x)}.
\label{eff solid}
\end{equation}
Here $m$ is the azimuthal order  , $k_{\rm b}$ is the flexural wavenumber of the corresponding
flat thin plate, defined by
$k_{\rm b}^2(\omega)=\omega\sqrt{\rho_d h/D}$, where
$\rho_d$, $h$, and $D$ are the material density, shell
thickness, and bending rigidity, respectively.

Equation~\eqref{eq:elastic_radial_wave} provides an elastic
counterpart to Eq.~\eqref{eq:radial_wave}: both describe
wave propagation and localization in an effective potential,
although their frequency dispersion differs.
The dependence $V_{\rm curv}\propto R^{-2}(x)$ allows the
shell profile to control confinement away from physical
boundaries. 

\section{Mapping 3+1D Spacetime into 2+1D Elastic shells}

To reproduce the radial dependence of the black-hole
photon-sphere potential, we construct a surface of revolution
for which $R^{-2}(x)$ encodes the factor $f(\rho)/\rho^2$
in Eq.~\eqref{eq:effective_potential}. The required relation
between the spacetime radial coordinate $\rho$ and the
shell profile follows from the Fermat metric\cite{frankel2011gravitational}, which preserves the spatial projections of null geodesics. Restricting to the equatorial plane, $\theta=\pi/2$, its spatial part reads
\begin{equation}
ds_F^2=f^{-2}(\rho)d\rho^2+\rho^2 f^{-1}(\rho)d\phi^2 .
\label{eq:fermat_metric}
\end{equation}
We identify this metric with that of a surface of revolution,
\begin{equation}
ds_{\rm shell}^2=dx^2+R^2(x)d\phi^2 ,
\end{equation}
where $x$ is the arclength along the meridional profile. Parametrizing the same surface by $\rho$ gives
\begin{equation}
dx=f^{-1}(\rho)d\rho,
\qquad
R(\rho)=\rho f^{-1/2}(\rho).
\label{eq:x_R_mapping}
\end{equation}
The second relation gives immediately
\begin{equation}
V_{\rm curv}(\rho)
=\frac{m^2}{ R^2(\rho)}
=\frac{m^2f(\rho)}{\rho^2},
\label{eq:curv_potential_mapped}
\end{equation}
which has the same radial dependence as the black-hole photon-sphere potential in Eq.~\eqref{eq:effective_potential}. The first relation fixes the corresponding meridional arclength and, together with $R(\rho)$, determines the embedded shell profile through \cite{xu2022light}
\begin{equation}
H(\rho)=
\int^\rho
\sqrt{
f^{-2}(\rho')
-
\left(\frac{dR}{d\rho'}\right)^2
}
\,d\rho' .
\label{eq:H_profile}
\end{equation}

\begin{figure}[!htbp]
\centering
    \begin{subfigure}{0.485\linewidth}
    	\centering	                
        \includegraphics[width=1\linewidth]{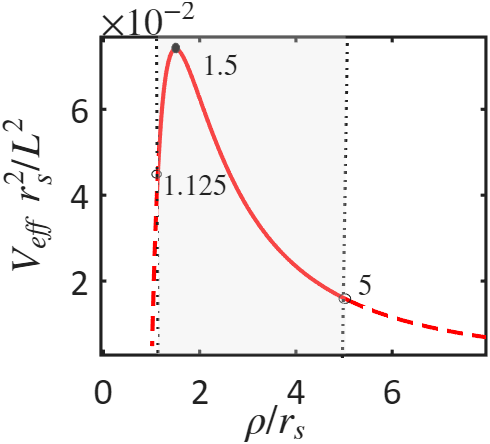}
	   \caption{}
         \label{poten SC}
    \end{subfigure}
    \hfill
    \begin{subfigure}{0.498\linewidth}
    	\centering	                
           \includegraphics[width=1\linewidth]{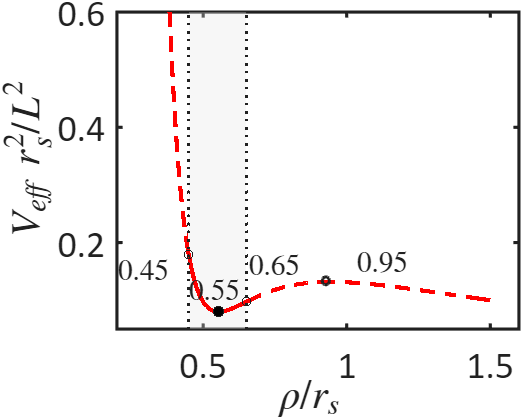}
       \caption{}
       \label{poten Rn}
    \end{subfigure}
    \caption{\protect\justifying 
    \textbf{Effective photon-sphere potentials as a function of the dimensionless radius \(\rho/r_s\).}
(a) Schwarzschild geometry, whose potential has a single maximum at \(\rho/r_s=1.5\), corresponding to the unstable photon-sphere orbit.
(b) Reissner--Nordstr\"om geometry with \(r_Q/r_s=0.512\), showing a stable orbit near \(\rho/r_s=0.555\).
The shaded areas delimit the radial intervals of the fabricated geometries.
}
    \label{fig:potentials}
\end{figure}

\begin{figure}[t]
    \centering
    \begin{subfigure}{0.48\columnwidth}
        \centering
        \includegraphics[width=\linewidth]{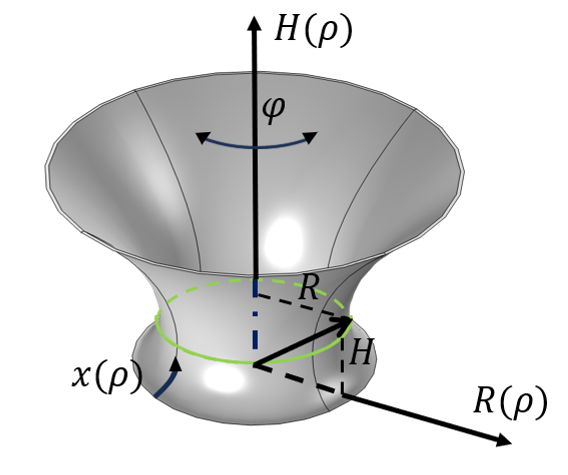}
        \caption{}
        \label{fig:schwarzschild_shell}
    \end{subfigure}
    \hfill
    \begin{subfigure}{0.48\columnwidth}
        \centering
        \includegraphics[width=0.9\linewidth]{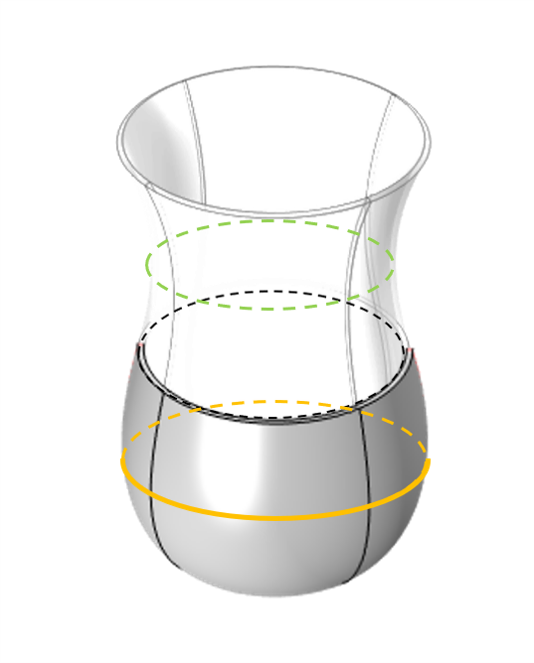}
        \caption{}
        \label{fig:rn_shell}
        \end{subfigure}
    \caption{\protect\justifying
    \textbf{Fermat-metric shell geometries.}
    (a) Schwarzschild analogue, showing the unstable photon-sphere orbit (green ring).
    (b) Reissner--Nordstr\"om analogue, showing the unstable photon-sphere orbit (green ring) and the stable photon-sphere orbit (yellow ring).
    The surfaces are generated from the profiles \(R(\rho)\) and \(H(\rho)\) obtained from Eqs.~\eqref{eq:x_R_mapping} and \eqref{eq:H_profile}. Both samples have uniform thickness.
    In panel~(b), the solid section denotes the fabricated shell used to probe the stable orbit, while the transparent extension shows the continuation of the same geometry toward the unstable orbit.
        }
\label{fig:shell_geometries}

\end{figure}

We apply this construction to two spherically symmetric geometries. The first is the Schwarzschild black hole, which is non-rotating and electrically neutral, with
\begin{equation}
f(\rho)=1-\frac{r_s}{\rho}.
\end{equation}
Its effective potential has a single maximum at $\rho=3r_s/2$, corresponding to an unstable photon sphere.

The second is the Reissner--Nordstr\"om geometry, describing a non-rotating charged black hole, for which
\begin{equation}
f(\rho)=1-\frac{r_s}{\rho}+\frac{r_Q^2}{\rho^2},
\end{equation}
where $r_Q^2=GQ^2/(4\pi\epsilon_0 c^4)$ is the charge length scale. In the superextremal range considered here, $0.5r_s<r_Q\lesssim0.53r_s$, the effective potential develops a local minimum, corresponding to a stable photon-sphere orbit. These two cases therefore provide complementary tests of curvature-induced confinement: an unstable photon sphere in the Schwarzschild analogue and a stable photon sphere in the Reissner--Nordstr\"om analogue.

The corresponding effective potentials are shown in Fig.~\ref{fig:potentials}.
The embedded shell geometries shown in Fig.~\ref{fig:shell_geometries}
are obtained by numerically evaluating \(R(\rho)\) and \(H(\rho)\) from
Eqs.~\eqref{eq:x_R_mapping} and \eqref{eq:H_profile}, and by revolving
the resulting meridional profiles around the symmetry axis.
These surfaces therefore provide finite elastic implementations of the
Schwarzschild and Reissner--Nordstr\"om geometries considered here.

The mapping therefore provides more than a geometrical
analogue of null trajectories: together with the elastic
wave equation, it yields a predictive description of the
photon-sphere mode spectrum. For a stable orbit at a local
maximum of the midsurface radius, $R_0=R(x_0)$, expanding
the effective potential about its minimum gives
\begin{equation}
f_{m,0}\simeq
\frac{B}{2\pi}
\left[
\frac{m^2}{R_0^2}
+|m|\sqrt{-\frac{R''(x_0)}{R_0^3}}
\right],
\label{eq:photon_sphere_frequency}
\end{equation}
where $B=\sqrt{D/(\rho_d h)}$ and the subscript $0$
denotes the meridional ground state. The first term is
the circular-orbit estimate, while the second accounts
for the finite meridional width of the localized mode.
The frequency is thus determined by the material
parameters, shell thickness, and local geometry,
without introducing an adjustable frequency offset.

Numerical comparisons show substantially improved
agreement in the frequency-versus-$m^2$ slope when
the shell thickness is reduced from $2$ to
$0.4\,\mathrm{mm}$,  [Fig. S1a and S1b] supporting the predicted scaling
in the thin-shell regime. A residual fitted frequency
offset remains unresolved. The derivation and the
thickness, truncation, and mesh comparisons are
presented in Sec.~B of the Supplemental Material.

\begin{figure*}[ht] 
    \centering              
    \includegraphics[width=0.9\linewidth]{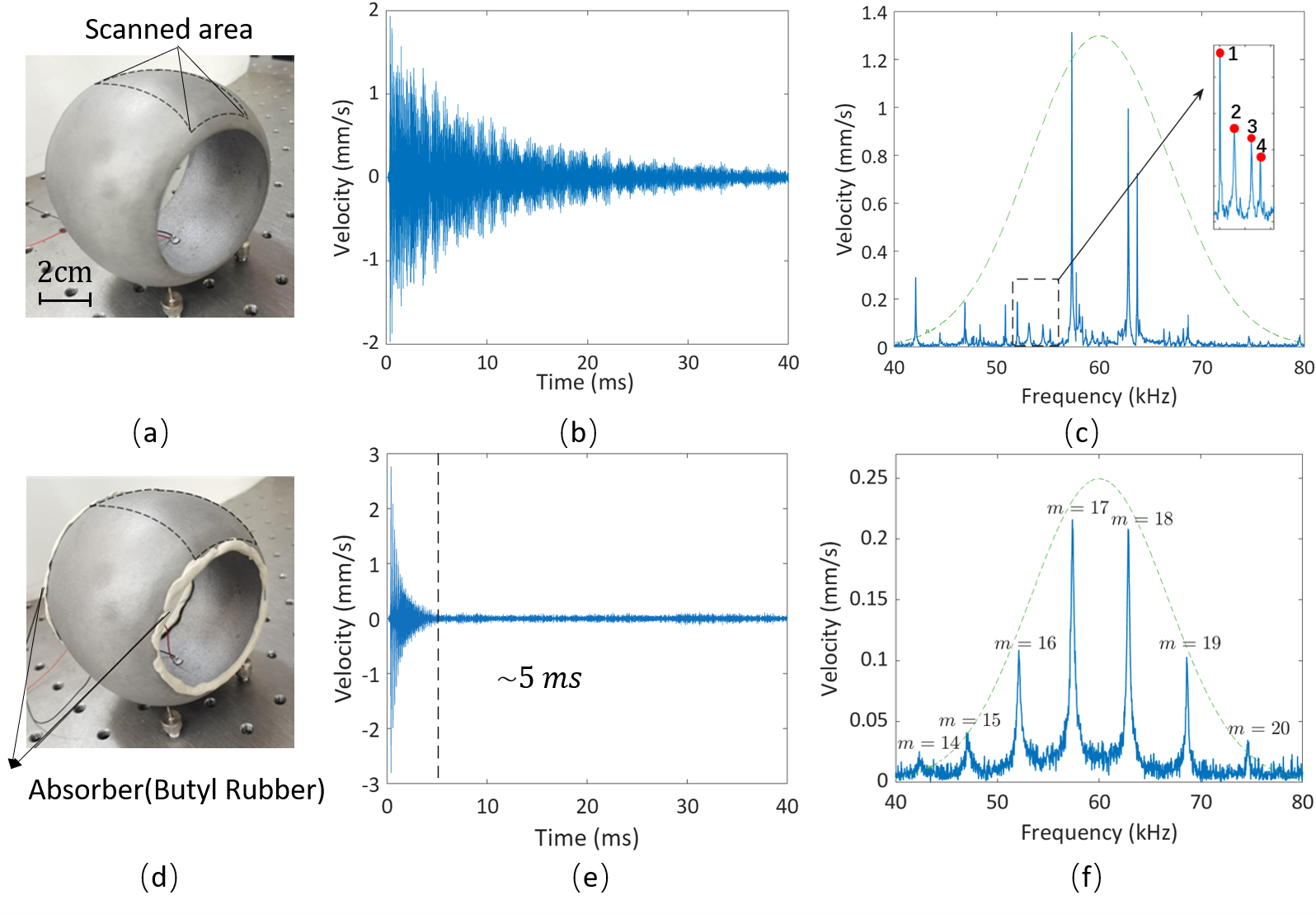} 
    \caption{\protect\justifying
    Results of the 3D surface measurements without and with absorbers. (a) and (d) shows the model with and without open boundaries; the dashed region indicates the scanned area, and the small white disk represents the transducer. (b) and (e) presents the averaged signal collected, while (c) and (f) displays the spectra of the scanned area. After introducing absorbers around the boundaries, the resonant peaks now differ solely in their azimuthal mode number $m$ (representing the number of nodes along the circumference). }
    \label{3D-absorber2} 
\end{figure*}

\begin{figure*}[!htbp] 
        \centering
        \includegraphics[width=0.9\linewidth]{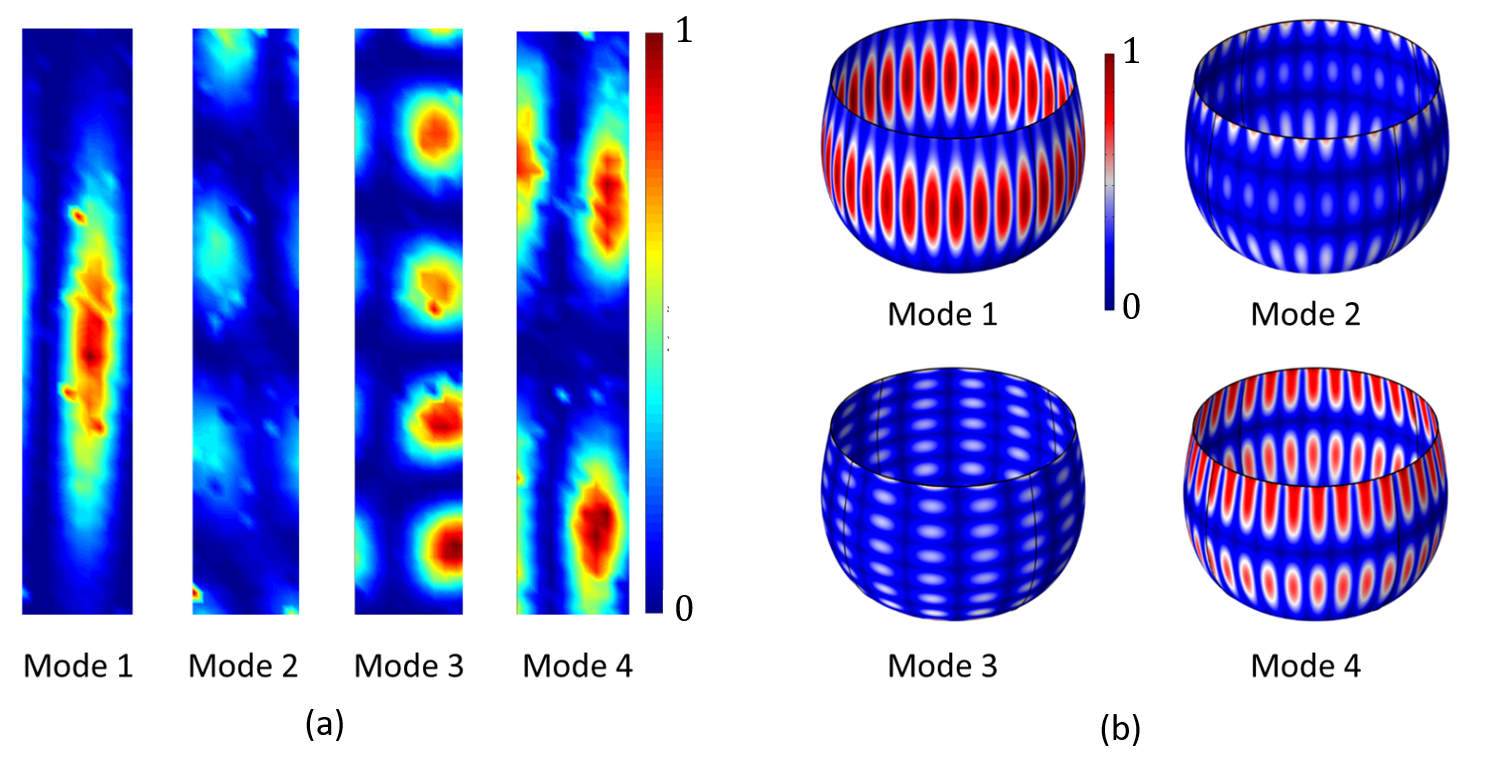}
        \caption{\protect\justifying
         (a) Measured mode projections onto a flat surface of width $10\,\text{mm}$ and height $50\,\text{mm}$. From left to right, the frequencies are 52.025 kHz, 53.15 kHz, 54.5 kHz, and 55.2 kHz, corresponding to modes 1, 2, 3, and 4 in Fig.~\ref{3D-absorber2}(c). The color scales are normalized to the maximum intensity(ranging from 0 to 1) of each individual mode for enhanced visualization of the modal structures;
      (b) Simulated full-mode patterns, the amplitude. The modes are highly symmetrical and show excellent agreement with the experimental measurements.}
        \label{modes-sim}
\end{figure*}

\section{Experimental implementation}

The two elastic analogues defined above are fabricated as three-dimensional AlSi10Mg aluminium-alloy shells by selective laser melting.
Both shells have a constant thickness of \(2~\mathrm{mm}\), measured along the local surface normal, so that flexural waves locally experience a plate of uniform thickness.
The Reissner--Nordstr\"om analogue implements the mapped interval \(0.45r_s \leq \rho \leq 0.65r_s\), centered on the stable photon-sphere orbit, and has a height of \(65~\mathrm{mm}\) and a maximum width of \(100~\mathrm{mm}\).
The Schwarzschild analogue extends from \(1.125r_s\) to \(5r_s\), with a height of \(48~\mathrm{mm}\) and a maximum radius of \(46~\mathrm{mm}\).
Each shell is mounted with its axis of revolution horizontal and supported by three small pins to minimize contact-induced losses, leaving the boundaries effectively free.
Flexural waves are launched by a \(5~\mathrm{mm}\)-diameter piezoelectric ceramic disk transducer bonded to the inner surface of the shell [Fig.~\ref{3D-absorber2}(a,d)].
The transducer is driven by \(60~\mathrm{kHz}\) sinusoidal voltage bursts with a Gaussian temporal envelope, generated by a computer-controlled waveform generator and amplified by \(20~\mathrm{dB}\). The resulting excitation has a typical spectral bandwidth of \(20~\mathrm{kHz}\) [Fig.~\ref{3D-absorber2}(c,f)].

The out-of-plane velocity field is measured on the outer surface by scanning laser Doppler vibrometry.
At each scan point, the time-resolved velocity signal is recorded and Fourier transformed, providing both the local ringdown response and the spatially resolved resonance spectrum. 

\section{Photon-Sphere Modes near the Stable Orbit of the Reissner--Nordstr\"om Shell}

We first consider the Reissner--Nordstr\"om shell shown in Fig.~\ref{3D-absorber2}a.
As defined in Fig.~\ref{fig:shell_geometries}b, this sample realizes the section of the analogue geometry centered on the stable minimum of the effective potential.

Following an impulsive excitation, the out-of-plane velocity field is recorded in the time domain at each scan point. The spatially-averaged time signal exhibits a clear ringdown lasting about \(40~\mathrm{ms}\), providing direct access to the relaxation dynamics of the elastic analogue , Fig.~\ref{3D-absorber2}b.

Fourier transforming the recorded traces yields the spatially averaged spectrum shown in Fig.~\ref{3D-absorber2}c, where several resonant peaks are observed.

Because the fabricated shell is necessarily truncated, its free-standing edges can act as reflective boundaries for elastic waves. They may therefore generate cavity-like resonances that depend on the particular finite size of the sample, rather than on the analogue black-hole potential. To test whether the observed peaks originate from such edge reflections, the free-standing edges are covered with absorbing material and the measurement is repeated. This procedure efficiently suppresses edge-supported flexural modes because they have large displacement amplitudes near the boundaries, where they strongly overlap with the absorber.
The temporal response then decays much faster, with a characteristic duration of about \(5~\mathrm{ms}\) [Fig.~\ref{3D-absorber2}e].
The spectrum is correspondingly simplified: most peaks disappear, while a small family of sharp resonances remains [Fig.~\ref{3D-absorber2}(f)]. The peaks removed by covering the free-standing edges are attributed to edge-reflection modes. The surviving resonances define a selected mode family whose spatial structure is analyzed next.

To resolve the spatial structure of the surviving resonances, we scan a \(10~\mathrm{mm}\times 50~\mathrm{mm}\) region of the outer surface with a spatial step of \(1~\mathrm{mm}\), and reconstruct mode maps from the complex Fourier amplitude of the measured velocity field at each resonance frequency. The scan provides the local out-of-plane response on a meridional section of the shell, while the full azimuthal pattern is reconstructed using the rotational symmetry of the structure. Representative maps are shown in Fig.~\ref{modes-sim}.
Modes~2--4 corresponding to peaks in the inset of Fig.~\ref{3D-absorber2}c, which are clearly visible for free-standing edges [Fig.~\ref{modes-sim}], disappear once the edges are covered with absorbing material. By contrast, the resonance labeled mode~1 remains well resolved after boundary absorption, together with higher-frequency resonances belonging to the same family. This different response to boundary absorption distinguishes this family from the edge-reflection modes. The corresponding spatial maps show that these resonances are localized away from the free-standing edges and concentrate around the region where the effective potential has its minimum.

\begin{figure}[!htbp]
    \centering
    \begin{subfigure}{0.49\linewidth}
        \centering
        \includegraphics[width=1\linewidth]{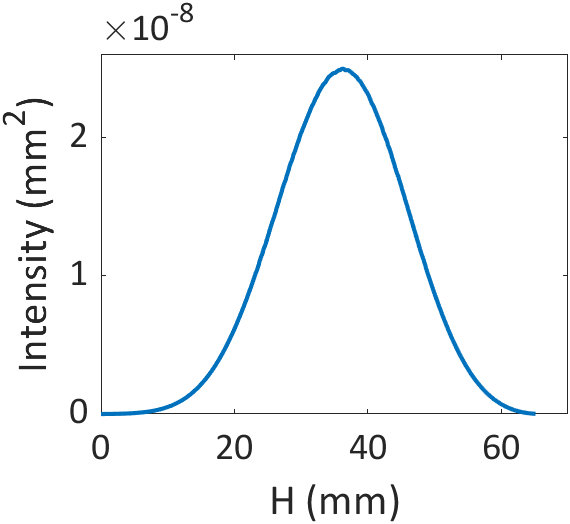}
        \caption{}
        \label{fig:PS-profile}
    \end{subfigure}
    \begin{subfigure}{0.49\linewidth}
        \centering
        \includegraphics[width=1\linewidth]{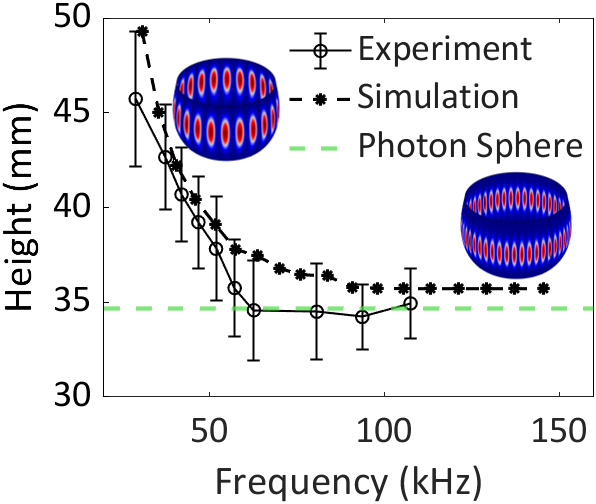}
        \caption{}
        \label{fig:PS-position}
    \end{subfigure}         
    \caption{\protect\justifying 
    (a) Profile of the mode at 70.466 kHz. (b) Frequency dependence of the position of the potential-confined modes. The mode position is defined by the axial coordinate $H_{\max}$ at which the velocity intensity reaches its maximum. Experimental and numerical results are shown as a function of frequency. The horizontal dashed line indicates the predicted stable photon-sphere position, $H_{\mathrm{PS}}=34.68~\mathrm{mm}$.}
    \label{fig:PS}
\end{figure}

To further characterize these modes and their spatial confinement, we first track the position of their maximum displacement as a function of frequency and determine whether it converges toward the predicted photon-sphere orbit. For the Reissner--Nordstr\"om geometry considered here, the stable photon-sphere orbit is predicted at $\rho_{\mathrm{PS}}=0.56,r_s$, corresponding in the fabricated shell to $H(\rho_{\mathrm{PS}})=34.68~\mathrm{mm}$ and $R(\rho_{\mathrm{PS}})=50.02~\mathrm{mm}$.

The positions extracted from experiment over the range $20$--$110~\mathrm{kHz}$ and from numerical simulations over $20$--$150~\mathrm{kHz}$ are reported in Fig.~\ref{fig:PS-position}. At low frequency, the displacement profiles are spatially extended and their maxima remain shifted from the geometrical photon-sphere position. As the frequency increases and the wavelength decreases, both the measured and simulated mode positions converge toward $H_{PS}=35\pm2~\mathrm{mm}$.
This frequency-dependent convergence shows that, in the short-wavelength regime, the surviving mode family becomes increasingly localized around the minimum of the effective potential, thereby identifying these resonances as photon-sphere modes.

\begin{figure}[!htbp]
    \centering
    \begin{subfigure}{0.49\linewidth}
        \centering
        \includegraphics[width=1\linewidth]{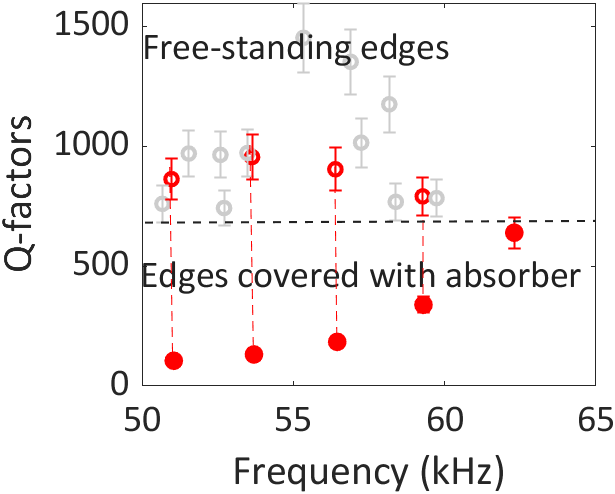}
        \caption{}
        \label{Q exper}
    \end{subfigure}
    \begin{subfigure}{0.49\linewidth}
        \centering
        \includegraphics[width=1\linewidth]{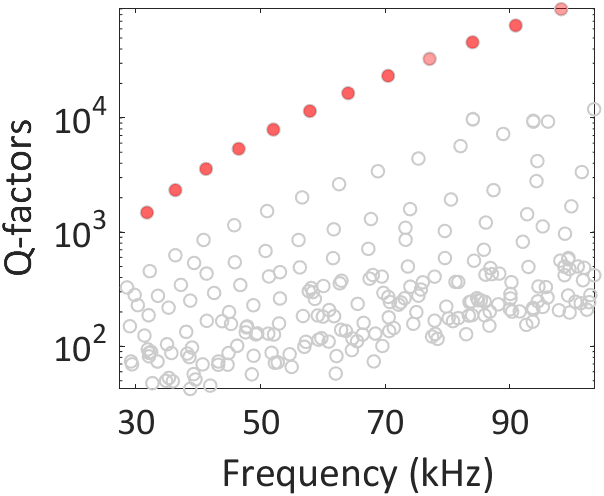}
        \caption{}
        \label{Q sim}
    \end{subfigure}  
    \caption{\protect\justifying
    (a) Experimental quality factors measured with free-standing edges and after the edges are covered with absorbing material. Open and filled symbols indicate the two boundary conditions, respectively. The photon-sphere modes are highlighted. (b) Numerical quality factors obtained from complex eigenfrequencies with perfectly matched layers at the shell boundaries. Photon-sphere modes form a distinct high-$Q$ branch.
    }
    \label{fig:Q-factor}
\end{figure}

We next characterize the confinement of the surviving mode family through its quality factor and its
sensitivity to boundary absorption. In the experiment, the quality factor is extracted from Lorentzian fits as $Q=f_r/\Delta f$, where $\Delta f$ is the full width at half maximum. In the numerical eigenfrequency analysis, the complex resonance frequency $f=f_r+i f_i$ gives $Q=f_r/(2|f_i|)$.

Figure~\ref{fig:Q-factor}a compares the experimental quality factors measured with free-standing edges and after the edges are covered with absorbing material. For the edge-sensitive resonances, the absorber produces a pronounced reduction of $Q$. By contrast, the potential-confined mode retains a comparatively large quality factor, showing that its decay is much less sensitive to the boundary conditions.

Figure~\ref{fig:Q-factor}b shows the numerical quality factors obtained from complex eigenfrequencies with perfectly matched layers at the shell boundaries. The photon-sphere modes form a distinct high-$Q$ branch, clearly separated from the remaining modes. Their quality factor increases with frequency as the modal profile becomes narrower and more strongly localized around the minimum of the curvature-induced effective potential.

The experimental quality factors remain systematically lower than the numerical values. This difference is attributed mainly to out-of-plane dissipation of the vibrating shell into the surrounding air, a loss channel that is not included in the present COMSOL simulations.
Overall, the weak response of the photon-sphere modes to boundary absorption and their distinct high-$Q$ behavior demonstrate that their confinement is governed primarily by the effective potential rather than by reflections from the finite shell edges.

Finally, to exclude any residual influence of the boundaries, we repeat the analysis using an extended Reissner--Nordstr\"om shell corresponding to the transparent continuation of the geometry shown in Fig.~2b. In this longer structure, the stable photon-sphere region is preserved, while the remote boundary is displaced substantially farther away.

The same photon-sphere mode family is recovered in the extended shell, with its intensity remaining localized around the minimum of the effective potential and with essentially unchanged spatial characteristics. By contrast, the cavity-like resonances are strongly modified by the increased system size and by the displacement of the boundaries. The complete experimental and numerical results are reported in the Supplemental Material.

The persistence of the photon-sphere mode under this substantial geometrical extension provides an independent test of its origin: its position and confinement are set by the curvature-induced potential, rather than by reflection at the truncated edges, imperfect absorption, or residual PML reflections. Together with the boundary-absorption measurements, this result rules out an interpretation in terms of conventional cavity or whispering-gallery modes. 

\begin{figure*}[!htbp]
    \centering
    \includegraphics[width=0.9\linewidth]{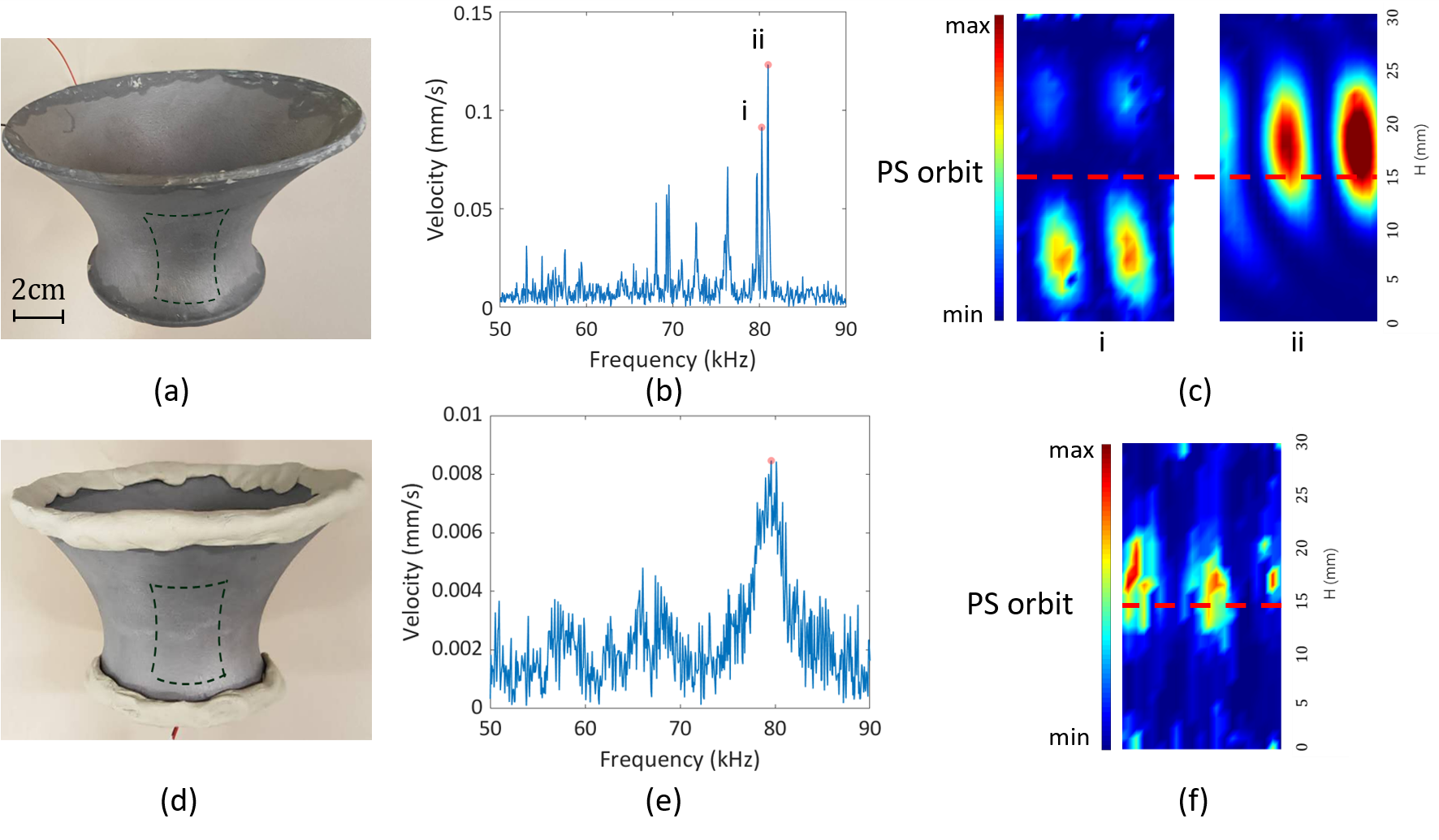}
    \caption{\raggedright
    Spatially resolved vibration measurements of the Schwarzschild
    shell with free-standing edges (a--c) and absorbing material
    applied to the boundaries (d--f).
    (a,d) Experimental configurations; the dashed regions indicate
    the scanned area.
    (b,e) Spectra obtained from the scanned area.
    (c,f) Planar projections of the measured mode patterns at the
    frequencies marked in red in (b) and (e), respectively.
    }
    \label{3D-absorber2-SCH}
\end{figure*}

\section{Photon-Sphere Modes near the Unstable Orbit of the Schwarzschild Shell}

We now turn to the Schwarzschild analogue, whose effective
potential has a single maximum at
$\rho_{\mathrm{PS}}=1.5\,r_s$, corresponding to an unstable
photon orbit. In the fabricated shell, this orbit lies at
approximately $H_{\mathrm{PS}}=14.5~\mathrm{mm}$. Unlike
the stable Reissner--Nordstr\"om case, this geometry provides
no potential minimum for stable trapping. Instead, we
investigate whether a leaky resonance can be observed near
the predicted potential maximum.

We follow the same experimental procedure as for the
Reissner--Nordstr\"om shell. Measurements on the Schwarzschild
geometry are more demanding because the rapidly varying
curvature makes it difficult to maintain the vibrometer
beam in focus across the entire surface. We therefore
restrict the scan to a $20~\mathrm{mm}\times30~\mathrm{mm}$
window centered on the predicted photon-sphere region
[Figs.~\ref{3D-absorber2-SCH}(a,d)].

With free-standing edges, several overlapping resonances
contribute to the measured response, producing multiple
spectral peaks and complex spatial patterns
[Figs.~\ref{3D-absorber2-SCH}(b,c)]. Applying absorbing
material to the boundaries strongly suppresses most of
these resonances and substantially shortens the ringdown.
Nevertheless, a weak but reproducible resonance remains
near $79.6~\mathrm{kHz}$
[Fig.~\ref{3D-absorber2-SCH}e].

The velocity map at this frequency displays a ring-like
pattern concentrated near $H\simeq16~\mathrm{mm}$
[Fig.~\ref{3D-absorber2-SCH}f], close to the predicted
photon-sphere position. Its survival under boundary
absorption, together with its concentration away from
the truncated edges, supports an interpretation in terms
of a curvature-controlled resonance rather than a mode
sustained primarily by reflections at the shell boundaries.

\begin{figure}[!htbp]
    \centering
    \begin{subfigure}{0.49\linewidth}
        \centering
        \includegraphics[width=\linewidth]{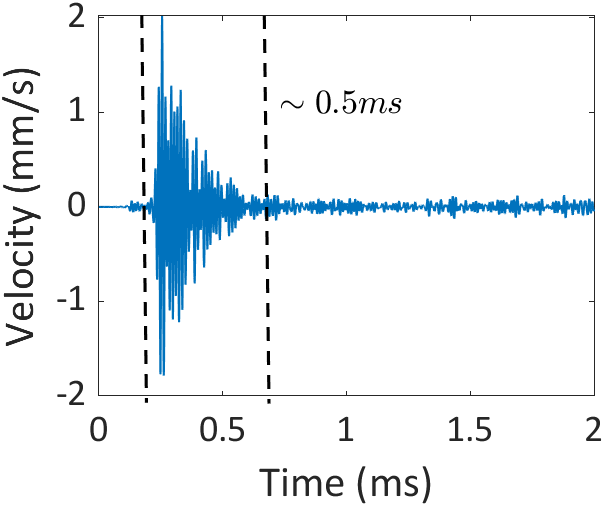}
        \caption{}
        \label{Singal-SCH}
    \end{subfigure}
    \hfill
    \begin{subfigure}{0.49\linewidth}
        \centering
        \includegraphics[width=\linewidth]{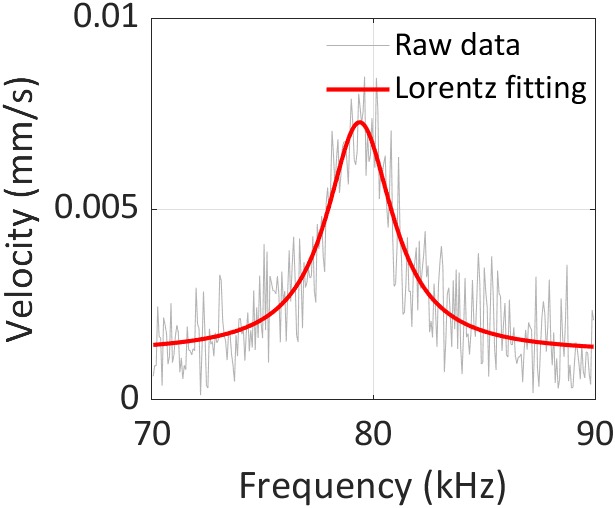}
        \caption{}
        \label{q-SCH}
    \end{subfigure}
    \medskip
    \begin{subfigure}[b]{0.49\linewidth}
        \centering
        \includegraphics[width=\linewidth]{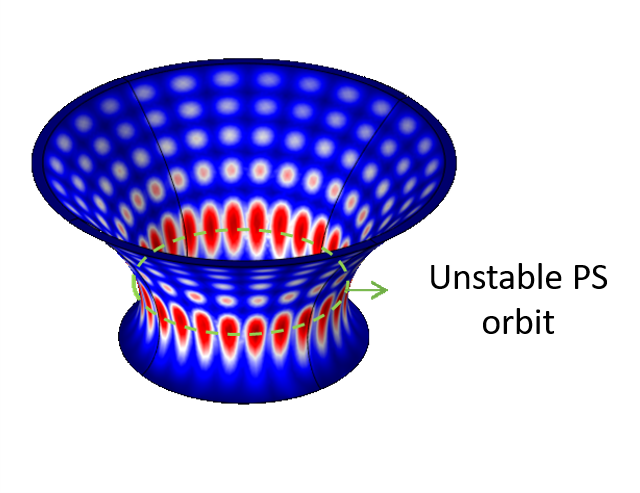}
        \caption{}
        \label{MODE-SCH}
    \end{subfigure}
    \hfill
    \begin{subfigure}[b]{0.49\linewidth}
        \centering
        \includegraphics[width=\linewidth]{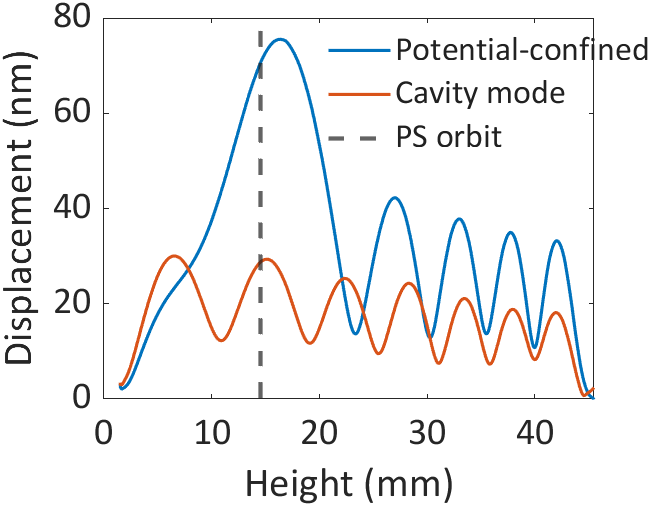}
        \caption{}
        \label{DIS-SCH}
    \end{subfigure}
    \caption{\raggedright
    Temporal, spectral, and spatial characteristics of the
    Schwarzschild-shell resonance.
    (a) Measured temporal response with absorbing boundaries,
    showing damped oscillations visible over approximately
    $0.5~\mathrm{ms}$.
    (b) Lorentzian fit to the surviving spectral peak near
    $79.6~\mathrm{kHz}$, yielding $Q=22\pm1.95$
    ($95\%$ confidence interval).
    (c) Simulated displacement-intensity distribution of a
    mode with a dominant ring near the predicted photon sphere.
    (d) Comparison of simulated intensity profiles for this
    mode and a boundary-confined mode, showing their distinct
    spatial distributions.
    }
    \label{Sch-Q AND S}
\end{figure}

The temporal response with absorbers exhibits damped
oscillations visible over approximately $0.5~\mathrm{ms}$
[Fig.~\ref{Singal-SCH}]. A Lorentzian fit to the surviving
spectral peak gives $Q=22\pm1.95$, with the uncertainty
corresponding to a $95\%$ confidence interval
[Fig.~\ref{q-SCH}]. These measurements resolve both the
spectral resonance and its transient decay, providing
a mechanical analogue of quasinormal ringing.

To further assess the spatial correspondence, we compare
the measured pattern with finite-element simulations of
the same geometry. The simulated mode exhibits a dominant
ring near the predicted photon sphere
[Fig.~\ref{MODE-SCH}]. Its intensity profile peaks in this
region, in contrast to the boundary-confined mode shown
for comparison [Fig.~\ref{DIS-SCH}]. The simulations thus
support the observed concentration near the potential
maximum. Because the experimental response is weak,
however, less intense features outside the dominant ring
may fall below the measurement noise floor.

Remarkably, the curvature-induced potential therefore supports a photon-sphere resonance even around an unstable orbit. This state is not a bound mode in the usual sense, but a leaky quasinormal resonance localized near the potential barrier and decaying through the open system. The analogy with black-hole ringing refers to this common modal structure and temporal relaxation, not the microscopic origin of the losses. In the elastic shell, dissipation arises from radiation into the surrounding air, material damping, and residual coupling to the supports, whereas in a black hole it is associated with gravitational radiation and absorption at the horizon. The Schwarzschild shell thus extends the
observation of curvature-controlled photon-sphere modes from stable trapping to the intrinsically radiative regime associated with an unstable photon orbit.

\section*{Conclusion}

We have realized a passive elastic analogue of black-hole photon-sphere dynamics and directly resolved the spectral, spatial, and temporal signatures of its quasinormal modes. By mapping Schwarzschild and Reissner--Nordstr\"om effective
potentials onto the geometry of curved metallic shells, we engineered flexural waves whose confinement is imposed by curvature rather than by conventional reflecting boundaries. The modes surviving at the predicted photon-sphere
orbits are therefore not ordinary eigenmodes of a finite shell, nor whispering-gallery modes sustained by repeated boundary reflection, but potential-induced resonances of an open wave system.

Beyond the geometrical mapping, we firstly show that the thin-shell reduction connects photon-sphere confinement to a quantitative frequency prediction, including the contribution of meridional localization. The improved slope agreement in thinner-shell simulations supports this description
and clarifies its range of applicability.

The elastic platform provides several decisive tests that were inaccessible in the previous optical realization. First, impulsive excitation gives direct access to the time domain and reveals a clear damped ringdown, together with the associated relaxation times. Second, scanning laser vibrometry maps the quasinormal-mode displacement field directly, without relying on gain or selective lasing. Third, Lorentzian fits to the resonances yield their quality factors and hence their decay rates, allowing the evolution of confinement and dissipation to be followed over a broad frequency range. Most importantly, covering the free-standing edges with strong absorbers suppresses the numerous boundary-supported resonances while leaving a distinct family of modes localized around the effective-potential extremum. Increasing the system size provides an independent test: the same photon-sphere families persist, including higher radial orders, while their localization remains tied to the potential rather than to the position of the truncation boundaries. Taken together, boundary absorption, size variation, direct mode mapping, frequency-dependent localization, and relaxation-time measurements establish a convergent body of evidence that these resonances are curvature-confined QNMs rather than WGMs of the fabricated shell.

A particularly unexpected result is that elastic confinement is observed not only around the stable Reissner--Nordstr\"om orbit, where a potential minimum naturally supports long-lived states, but also near the unstable Schwarzschild photon sphere. This provides a laboratory view of how a wave can transiently concentrate around an unstable orbit and subsequently decay, thereby connecting photon-sphere localization to ringdown in a directly observable mechanical system. The analogy concerns the effective potential, open-system resonance structure, and damped temporal response; the microscopic loss channels are, of course, platform dependent. In the elastic shell, decay includes radiation into the surrounding air, material damping, support losses, and residual boundary leakage, whereas black-hole QNMs decay through gravitational radiation and absorption by the horizon. Distinguishing these channels will be essential for quantitative comparisons of imaginary eigenfrequencies.

Beyond analogue gravity, these results identify curvature as an independent design parameter for controlling localization, lifetime, and radiation in elastic structures. The present approach can be extended to rotating
Kerr-inspired geometries, coupled or multiple photon spheres, deliberately introduced inner reflectors for analogue echoes, non-Hermitian and nonlinear regimes, and planar implementations obtained through transformation elastodynamics. It also suggests mechanically robust resonators whose modes are less sensitive to edge imperfections than boundary-confined states. More fundamentally, the ability to excite, image, and follow the decay of individual photon-sphere modes turns black-hole ringdown from a remotely inferred astrophysical signal into a spatially and temporally resolved laboratory wave phenomenon.


\textbf{Acknowledgments -} This work was supported by The Israel Science Foundation grants 1871/15, 2074/15, 2630/20, 1698/22 and 3777/25; the United States–Israel Binational Science Foundation NSF/BSF Grant 2015694 and 2020245, and the BSF Grant 2022158. P.S. acknowledges the precious help of his lab manager, Dr. Leonid Wolfson. J.W. has been supported by the Excel@BINA scholarship from the Nanocenter at Bar Ilan University.

\bibliographystyle{apsrev4-2}
\bibliography{Reference}

\end{document}